\documentclass[twocolumn,letterpaper,10pt]{article}
\usepackage[dvips]{graphicx}
\usepackage{epsfig}
\usepackage{boxedminipage}
\usepackage{subfigure}
\usepackage{caption}
\usepackage{longtable}
\usepackage{amsmath}
\usepackage{amssymb} 
\usepackage{euscript}
\usepackage{subfigure}
\usepackage{appendix} 
\usepackage{lineno} 
\usepackage{hyperref}
\usepackage{breakurl} 

\usepackage{times}		
\hypersetup{
  pdfauthor = {Martin Leitgab},
  pdftitle = {SSP Design Competition Entry},
  pdfkeywords = {space solar power, wireless power transmission }
}

\usepackage{titlesec} 
\titlespacing\section{0pt}{12pt plus 0pt minus 0pt}{4pt plus 0pt minus 0pt}
\titlespacing\subsection{0pt}{12pt plus 0pt minus 0pt}{4pt plus 0pt minus 0pt}
\titleformat{\section}{\bfseries}{\thesection}{1em}{}
\titleformat{\subsection}{\bfseries}{\thesubsection}{1em}{}
\begin{document}
\urlstyle{rm}
\captionsetup{margin=7pt,font=footnotesize,labelfont=bf}

\twocolumn[
  \begin{@twocolumnfalse}
  \begin{flushleft}
    {\large $2^{\textnormal{nd}}$ Space Solar Power International Student and Young Professional Design Competition} \\
\vspace*{3.0mm}
    \textbf{\LARGE Hypermodular Distributed Solar Power Satellites - \\ 
    \vspace*{1.5mm}
    Exploring a Technology Option for Near-Term LEO \\ 
		    \vspace*{0.5mm}
				Demonstration and GLPO Full-Scale Plants} \\ 
  \end{flushleft}
  \vspace*{1mm}
  \begin{flushleft}
    \textbf{\large Martin Leitgab, PhD}\\
  \vspace*{1.5mm}
    \textit{Physics Department, University of Illinois at Urbana-Champaign, 1110 West Green Street, Urbana, IL 61801, USA \\
      leitgabm@gmail.com} \\ 
  \end{flushleft}
\begin{abstract}
This paper presents a new and innovative design for scaleable space solar power systems based on satellite self-assembly and microwave spatial power combination. Lower system cost of utility-scale space solar power is achieved by independence of yet-to-be-built in-space assembly and transportation infrastructure. Using current and expected near-term technology, this study explores a design for near-term space solar power low-Earth orbit demonstrators and for mid-term utility-scale power plants in geosynchronous Laplace plane orbits. High-level economic considerations in the context of current and expected future launch costs are given as well. 
\end{abstract}
  \vspace*{5mm}
  \end{@twocolumnfalse}
  ]
  \parskip 0pt

\section{Introduction}
\label{sec:intro}

\subsection{Space Solar Power}

The global energy demand is estimated to increase by $30\%$ until $2035$~\cite{ieareport} and to continue to grow thereafter. Fossil fuels are non-renewable resources and will likely not be able to meet the increasing energy demand towards the end of the current century. Energy production from fossil fuels in general causes the emission of greenhouse gases which are connected to man-made climate change~\cite{ipcc4threportwg1}. Space solar power (SSP), first proposed in Ref.~\cite{glaser}, represents a large-scale renewable energy source with minimal strain on the environment. Several system-level studies have concluded that SSP is technically feasible (e.g., the $2011$ IAA study, Ref.~\cite{iaastudy2011}). Numerous proof-of-principle technology experiments were conducted (for an overview see Refs.~\cite{iaastudy2011} and~\cite{leopoldpaper}). More recently, advanced analyses of the financial viability of SSP were performed and found promising results (e.g., Ref.~\cite{mankinsniac}). Ongoing system integration studies to produce a subsystem SSP prototype are performed at the Naval Research Laboratory, USA~\cite{jaffepaper}. 


This paper combines recent experimental SSP system integration and theoretical orbit selection results with aspects of previous SSP system architecture work to formulate a scalable SSP design independent of in-space assembly and transportation infrastructure. These auxiliary systems would need to be developed in parallel to the main SSP systems, require additional resources and might prevent realization of utility-scale SSP as a whole due to the large total amount of required resources. Time to deployment of the first utility-scale SPS systems is expected to be shortened if the SSP architecture does not rely on in-space assembly or transportation infrastructure. The goal of the pathway illustrated in this paper is to realize utility-scale SSP in the mid-term.

\subsection{Hypermodular Distributed Space Solar Power (HD-SSP)}

Hypermodularity is an attractive principle for space solar power architecture, as e.g. recently used in Ref.~\cite{mankinsniac}. By assembling full SSP stations from a large amount of identical elements, production cost are significantly decreased and reliability of the system is increased due to fewer single points of failure. 

On the other hand, the concept of distributed wireless power transmission by means of spatial microwave power beam combination has not received much usage in most recent SSP system designs. Spatial beam combination aims at combining radiation beams from multiple sources at arbitrary (but not too far) locations with respect to each other into one coherent beam at the ground receiver site. Beam combination as a concept for SSP has already been discussed in previous research on a conceptual level, e.g. Ref.~\cite{Yakovlevpaper}. However, the author is unaware of major recent research endeavors in this field. A general complication of using spatial beam combination from independent transmitters for SSP is the 'thinned-array-curse', described in Ref.~\cite{wikipediathinnedarraycurse}. This idea relates empty space between independent radiators to a loss of received power proportional to the inter-radiator distance, for a given rectenna on the ground. The present study explores system designs with the attempt to avoid such losses for distributed wireless power systems based on free-flying power radiators.

The SSP system design presented in this paper bases heavily on hypermodularity and distributed wireless power transfer. The concept is therefore named hypermodular distributed space solar power (HD-SSP). Even though spatial beam combination represents the aspect of HD-SSP with the lowest technological readiness level, current off-the-shelf technology could be of sufficient precision to enable fast research and development for near- to mid-term implementation of this concept. Industrial laser-based tracking systems measure distances of around $50$~m to a precision of better than $3$~permille of the $2.45$~GHz wavelength~\cite{lasertrackerpaper}. This precision could be enough for appropriate choice of beam phase and direction for each HD-SPS to achieve high beam combination efficiency at the receiver site. Additional research and development would be required to explore this option for SSP architectures.

All calculated results given in this paper are rounded to two significant digits.


\section{HD-SSP LEO Demonstrator}

Multiple options for electricity generation and wireless power transmission have been considered in the history of SSP. For simplicity, the presented design focuses on photovoltaic cells and microwave power transmission at $2.45$~GHz using solid state power amplifiers. Optimization of the design with different subsystem choices can be performed at a later stage. In the following, the system layout of a hypermodular distributed solar power satellite (HD-SPS) is described.

\subsection{General HD-SPS System Layout}

Each HD-SPS consists of a main platform composed of several hexagonally-shaped 'sandwich' structures similar in design to the 'hexbus' sandwich structures proposed in Ref.~\cite{mankinsniac}). The sandwich concept has been introduced to facilitate power management and distribution (PMAD) by locating the power generation and power transmission surfaces of a SPS close to each other in a parallel, sandwich-like layering. Generic satellite systems, such as thermal management systems, guidance, navigation and control, command and data systems, and communication systems, are also part of the design of the main platform. 

Due to orbital perturbations in LEO such as atmospheric drag and gravitational perturbations, station-keeping systems are required to maintain orbit. For a mid-term, utility-scale realization of the HD-SPS concept, electric propulsion seems the most promising technology choice. Power to the propulsion units could be delivered either through the main SPS PMAD system or through an independent circuit from additional, dedicated photovoltaic modules. The latter choice would probably reduce the complexity of the main sunlight-to-RF PMAD system and further contribute to the modularity of the concept. If development of the electric propulsion system cannot be completed in time, LEO demonstrator SPS might still need to be flown with conventional, chemical propulsion thrusters. 

For SPS architectures in which the power transmission surface of the sandwich modules is to be pointed towards earth at all times, a mirror/reflector structure is required to provide illumination on the power generation surface throughout the entire orbit. Therefore HD-SPS feature a reflector structure consisting of individually controlled reflector elements, movable about one or two axes. The reflector array could also serve as power input level conditioning if it is designed to be capable of shielding parts or all of the power generation surface from sunlight. 

One of the basic design principles of HD-SPS is independence of in-space assembly infrastructure. Thus each unit will self-assemble on orbit. This can be achieved, e.g., through spring-loaded interconnects between sandwich modules which will unfold and deploy the sandwich platform once released in microgravity. Other deployment options include one-time-use electric motors on each sandwich module which are activated as soon as the respective module is exposed to sunlight. The reflector array can be deployed via inflation or roll-out mechanisms. 


\subsection{Sizing and Power Scales of Near-Term HD-SPS}

To facilitate self-assembly on orbit, the size of a complete HD-SPS system is chosen to match the payload volume and payload mass of current launch vehicles to LEO, such as the Falcon $9$ launch vehicle of SpaceX~\cite{spacexfalcon9}. The Falcon$9$ vehicle provides a cylindrical volume for payload with about $4.6$~m diameter and $6.6$~m height, in addition to a conically-shaped volume with a maximum diameter of $4.6$~m, a minimum diameter of $1.3$~m and a height of $4.8$~m. For simplicity, it is assumed that the cylindrical volume can be filled entirely with flat sandwich structures stacked along the center axis of the launch vehicle, while the conical volume is filled with the reflector array and the remaining satellite subsystems. In the calculation of the results mentioned in the following, an effective sandwich payload volume is assumed as a cylinder with $4.4$~m diameter and a height of $6.5$~m. This allows $10$~cm for additional protective packing material against shock and vibrations on the sides and bottom of the volume, corresponding to a volume reduction of about $10\%$. 

The performance of the sandwich modules in a near-term SSP demonstrator will likely be close to the performance achieved for sandwich modules in current research. Reference~\cite{jaffepaper} recently reported an area-specific mass of about $19$~kg/m$^{2}$ for a prototype sandwich module, however without essential power transmission electronic elements such as phase shifters. The end-to-end efficiency from incident sunlight to radio-frequency (RF) power (with simulated antenna efficiency) is reported to be about $12\%$. For the present study, an additional $1$~kg/m$^{2}$ of material is allotted for RF power conditioning devices, raising the area-specific mass to $20$~kg/m$^{2}$. The total prototype height in Ref.~\cite{jaffepaper} is about $12$~cm~\cite{perscorrJaffe}. It is assumed that sandwich modules similar to the prototype would consume $15$~cm of effective height inside a launch vehicle, which includes protective packing material against shock and vibrations.

Table~\ref{tab:payloadconfigsmassvsvol} shows sandwich module specifications, number of sandwich modules stacked along vertical direction in the cylindrical Falcon $9$ payload volume and approximate nameplate RF power per Falcon $9$ launch, assuming that $90\%$ of the total surface area is active photovoltaic area. The first and second column quote quantities for achieved sandwich performance parameters for square and hexagonal sandwich shape, respectively. It can be seen that the hexagonal base shape achieves about $30\%$ higher RF power than the choice of square-shaped modules, but also produces a $30\%$ heavier total mass.

The third and fourth column in Table~\ref{tab:payloadconfigsmassvsvol} give quantities for square and hexagonal module shape under estimation of a near-term $20\%$ reduction of area-specific mass and effective module height, which is expected to be achievable by a dedicated, moderately-funded system integration effort. The last column additionally assumes a $20\%$ relative increase of total sunlight-to-RF sandwich efficiency from currently achieved $12\%$ to about $15\%$ for hexagonal module shape. With such sandwich module specifications, a single Falcon $9$ launch could deliver a single HD-SPS to LEO with a nameplate RF power level of $120$~kW at one-sun concentration. This case is used for the remainder of this paper as a reference design for a HD-SPS LEO demonstrator. The estimated mass of the sandwich modules in this design is similar to the 'DRM\_$1$' scenario of a LEO demonstrator in Ref.~\cite{mankinsniac}, which however projects lower RF output.  

	\begin{table*}[t]
	\centering
		\begin{tabular}{|c|c|c|c|c|c|} \hline
		          & Square        & Hexagonal       & Improv. Square& Improv. Hex.  & Improv.$^{2}$ Hex \\ \hline
		Mod. Area & $10$m$^{2}$   & $13$m$^{2}$     & $10$m$^{2}$   & $13$m$^{2}$   & $13$m$^{2}$   \\ \hline
		Mod. Mass & $200$~kg      & $250$~kg        & $160$~kg      & $200$~kg      & $200$~kg      \\ \hline
		Nr. Mod.  & $43$          & $43$            & $53$          & $53$          & $53$          \\ \hline
		Tot. Mass & $8.3$~tons    & $10.8$~tons      & $8.3$~tons    & $10.8$~tons    & $10.8$~tons    \\ \hline
		Tot. Area & $410$~m$^{2}$ & $540$~m$^{2}$   & $520$~m$^{2}$ & $670$~m$^{2}$ & $670$~m$^{2}$ \\ \hline
		RF Power  & $62$~kW       & $81$~kW         & $78$~kW       & $100$~kW       & $120$~kW       \\ \hline
				\end{tabular}
		\caption{Calculations of expected single HD-SPS nameplate RF power using experimental (Ref.~\cite{jaffepaper}) and estimated near-term sandwich performance parameters together with SpaceX Falcon $9$ payload volume specifications. The assumed incident power at one-sun concentration is about $1400$~W/m$^{2}$. Further description is given in the text.}
\label{tab:payloadconfigsmassvsvol}
	\end{table*}

\begin{figure}[h]
	\begin{center}
		\includegraphics[width=0.2\textwidth]{{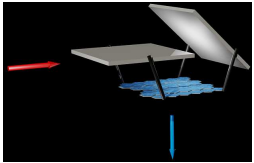}} 
		\caption{\label{fig:leodemo} 
		A possible realization of a LEO HD-SPS demonstrator with sandwich platform parameters as specified in the HD-SPS LEO reference design (parameters in right-most column of Table~\ref{tab:payloadconfigsmassvsvol}). The demonstrator is shown passing over the receiver site at dawn. The view direction of the drawing is south along the Earth's north-south axis. The red arrow indicates the direction of incoming sunlight, the blue arrow depicts the direction of the RF beam.}  
	\end{center}
\end{figure}

Figure~\ref{fig:leodemo} illustrates a potential assembly of a HD-SPS LEO demonstrator in the context of the HD-SPS LEO reference design, with $41$ sandwich modules arranged in a square shape with about $26$~m side length. The reflector array is reduced to very elementary features and is of mere symbolic character.

\subsection{Launch Vehicle Payload Capacity Considerations for a LEO HD-SPS Demonstrator}

The HD-SPS LEO demonstrator reference design uses about $69\%$ of the payload volume in a Falcon $9$ launch vehicle for the sandwich modules of the LEO demonstrators. It is assumed that generic satellite systems and the reflector array can be stowed in the remaining $31\%$ of the payload volume. More critically, sandwich modules occupy $82\%$ of the payload mass in the reference scenario, leaving about $2.4$~tons of payload mass available for generic systems and the reflector array. Assuming $5\%$ of the total payload mass being used for protective material during launch, and considering the estimated weight of the reflector array of about $500$~kg, other systems are not expected to be adding substantial mass beyond about $1.3$~tons. Therefore the HD-SPS LEO demonstrator reference design is considered a realistic near-term LEO demonstrator architecture. 

\subsection{LEO HD-SPS Ground Systems}

In order to demonstrate full SSP functionality, also a ground station is part of the HD-SPS LEO reference design. The ground station consists of a rectenna to convert the intercepted RF power from the demonstrator to electricity. In addition, it also produces the pilot signal required for retrodirective RF beam pointing and phase control and acts as a command and control center to provide monitoring of the satellites and the microwave beams. To save cost, the rectenna for the LEO demonstrator could be built at the same location as the rectenna for a mid-term utility-scale constellation of HD-SPS. Considering a utility-scale HD-SPS constellation providing $1$~GW of RF power described in Section~\ref{sec:glpoconstellation}, the rectenna diameter to catch $95\%$ of the transmitted RF power according to Ref.~\cite{jaffepaper},

\begin{equation}
\tau = \frac{\sqrt{A_{t}A_{r}}}{\lambda D},
\label{eq:taueq}
\end{equation}

is equal to about $7.8$~km. A LEO demonstrator irradiating the same rectenna could be deployed up to orbital altitudes of $750$~km while still achieving a power coupling efficiency of $95\%$. If the LEO demonstrator is launched to an orbital altitude of only $200$~km, the rectenna size could be decreased to $1.8$~km diameter. 

To control power flow from a SPS ground station into the local electricity grid, a fuel production plant can be attached to each rectenna, as described in Ref.~\cite{sneadpaper}, e.g. plants producing hydrogen through electrolysis of water. In times of low grid power demand, parts or all of the received power of HD-SPS can be used to produce fuel. Other options for a 'dispatchable load' between the rectenna and the local electricity grid include water desalination plants.

\subsection{Cost Estimates for LEO Demonstration HD-SSP}
\label{sec:leodemocost}
An order-of-magnitude cost estimation analysis for two LEO demonstrators (to demonstrate power beam combination) and a corresponding ground rectenna system will be performed. Ref.~\cite{IanRadicipaper} quotes average production cost for communication satellites around $200,000$~USD/kg. It is assumed that this value includes cost for research and development, test and evaluation (RDT$\&$E) and launch. In the presented study, production cost will be estimated separately from development and launch cost. Thus an initial production cost of only $100,000$~USD/kg is used.

The model for estimating an order-of-magnitude cost for developing, manufacturing and deploying two HD-SPS LEO demonstrators is based on the cost model described in Ref.~\cite{IanRadicipaper}. This model uses the learning curve method, where total production cost $C_{\textnormal{tot}}$ of a number $N$ of identical units is estimated as $C_{\textnormal{tot}} = C_{\textnormal{first}} \times N^{B}$, where the parameter $B$ is calculated from the assumed slope of the learning curve $k_{lc}$ as $B = 1- ln(1/k_{lc})/ln(2)$. In the presented study, slopes $k_{lc}=0.95$ for $N<10$ and $k_{lc}=0.90$ for $10<=N<50$ are adopted from Ref.~\cite{IanRadicipaper}, whereas the slope value $k_{lc}=0.85$ for $N>=50$ is taken from Ref.~\cite{Jillathesis}. 

For the HD-SPS LEO reference model parameters given in the last column of Table~\ref{tab:payloadconfigsmassvsvol}, the sandwich module mass of $200$~kg results in production cost for the first demonstrator sandwich of about $20$~million USD. Applying the learning curve method for producing the about $110$ identical sandwich modules required for two LEO demonstrators yields a total sandwich production cost of about $730$~million USD. Ref.~\cite{IanRadicipaper} estimates non-recurring RDT$\&$E cost by multiplying the cost of the first unit with a factor of $5.5$, for the case of the sandwich modules resulting in about $110$~million USD. 

Additional generic satellite systems such as guidance, navigation and control, command and data systems and communication systems are expected to be established technology without the need of large amounts of research and development. It is assumed that pure production costs for generic subsystems can be estimated by $5\%$ of the total sandwich production cost for one demonstrator. This cost is extrapolated for two demonstrator satellites using a learning curve slope of $0.95$, yielding a total production cost for generic subsystems of about $56$~million~USD. Conservatively, still a non-recurring generic subsystem RDT$\&$E cost contribution of $5.5$ times the cost of the initial subsystems is assumed, which yields a cost of $160$~million~USD.

The mirror elements, support structure with light-weight tethers and mirror attitude control in the HD-SPS reflector array have to be developed and integrated. Total RDT$\&$E and production cost of the reflector array are estimated the same way as for generic subsystems above. In addition, the concept of HD-SPS involves self-deployment of all sandwich modules and the reflector array once the final orbit is reached. Payload integration into the launch vehicle and deployment mechanisms and concepts have to be developed and tested. Another analogous amount for RDT$\&$E and production costs as for generic subsystems and the reflector array are added to cover these expenses.

SpaceX quotes $54$ million USD as cost for launch of a Falcon$9$ vehicle with about $13$~tons of payload to LEO~\cite{spacexfalcon9} (corresponding to a specific launch cost of about $4100$~USD/kg). Two launches will be performed to deploy two demonstrator HD-SPS satellites.

The rectenna size and thus cost depends on the chosen orbit which the demonstrators are deployed to. For an HD-SPS LEO demonstrator orbit of $300$~km altitude, a $3.2$~km diameter rectenna is required to intercept $95\%$ of the power. Assuming the same cost of about $10$~USD/m$^{2}$ as used in Ref.~\cite{mankinsniac}, the total rectenna cost contribution for LEO demonstrators is about $80$~million USD. On the other hand, if the LEO demonstrator is launched to an orbital altitude of only $200$~km, the rectenna size could be decreased to $1.8$~km diameter, corresponding to a cost contribution of about $27$~million USD. A trade study will have to be performed to find the most economical orbit altitude in terms of propellant required for orbital transfer from launch to final orbit, propellant required for station-keeping due to perturbations such as atmospheric drag, and rectenna cost from the area required to intercept $95\%$ of the transmitted power. For the exemplary $300$~km altitude orbit, a factor of $3$ times the rectenna cost is assumed to cover cost for ground-based rectenna RDT$\&$E, amounting to about $240$~million USD. 

Table~\ref{tab:leoproductioncost} summarizes the calculated cost contributions from development, production and launch of two HD-SPS satellites in addition to costs for a rectenna ground station. The total cost of a SSP demonstration system consisting of two HD-SPS satellites in an exemplary $300$~km orbit altitude is estimated to be around $1.9$~billion USD.

	\begin{table*}[t]
	\centering
		\begin{tabular}{|c|c|} \hline
	                 {\bf Item}      & {\bf Est. Cost} \\ \hline
Sandwich Production + RDT$\&$E     & $730+110$~M~USD \\ \hline
Generic Satellite Systems          & $56+160$~M~USD \\ \hline
Reflector Array                    & $56+160$~M~USD \\ \hline
Vehicle Integ., Deployment Concept & $56+160$~M~USD \\ \hline
Launch                             & $54+54$~M~USD \\ \hline
Rectenna                           & $80+240$~M~USD \\ \hline
{\bf Total }                       & $1.9$~B~USD \\ \hline
				\end{tabular}
		\caption{Estimated total cost for RDT$\&$E, production and launch of the first two LEO HD-SPS demonstrators, launched to $300$~km orbit altitude. Included are estimated production/installation and RDT$\&$E costs for a ground station rectenna to intercept $95\%$ of the transmitted RF power.}
\label{tab:leoproductioncost}
	\end{table*}

The cost contributions derived in this Section should be understood as rough order-of-magnitude estimations of potential costs to develop and deploy HD-SPS LEO demonstrators. The analysis does not claim to be complete or comprehensive.


\section{Mid-Term HD-SPS Constellations in Geosynchronous Orbit for Utility-Scale Baseload Power}

As mentioned in the introduction, the design concept of HD-SSP bases on hypermodularity and distributed wireless power transfer. In the previous Sections, the system layout for near-term LEO demonstration HD-SSP satellites has been described. The concept of HD-SSP is scaleable to utility RF power levels by combining the microwave beams of many satellites which are deployed to geosynchronous orbits but which are otherwise similar to the LEO demonstration reference systems, as will be described in the following.

\subsection{Orbit Selection- Geosynchronous Laplace Plane Orbits}

Recent analyses of geosynchronous SPS orbits alternative to GEO revealed that for certain orbits perturbing forces on satellites mostly cancel~\cite{Ianisdc13}. An example for such orbits are geosynchronous Laplace plane orbits (GLPOs). The assumption for HD-SPS is that once in GLPO, only vanishing amounts of propellant have to be used for station-keeping. The only relevant propellant need is given for 3-axis stabilization the satellites to maintain the RF sandwich plane area vector parallel to the ground rectenna area vector. Since GLP orbits are essentially 'frozen', also constellations and relative positions of HD-SPS within a constellation are assumed to remain virtually unperturbed. This eliminates the otherwise required propellant mass ballast even after orbital transfer from LEO to GLPO is completed.

\subsection{GLPO HD-SPS System Layout}
\label{sec:glpoconstellation}
Moderate mid-term technological progress is assumed to occur between the production of LEO HD-SPS demonstrators and GLPO HD-SPS for utility-scale baseload power production. This is expected to result in a $40\%$ reduction of currently achieved area-specific mass to about $12$~kg/m$^2$ and a reduction of effective sandwich height, including protective packaging, to about $9$~cm. In addition, a $40\%$ increase in total sandwich module efficiency to about $17\%$ is assumed. Finally, the thermal properties of mid-term HD-SPS are expected to allow three-sun concentration on the photovoltaic surface of the sandwich modules. Three-sun concentration is quoted as achievable for existing sandwich prototypes in Reference~\cite{jaffepaper} with moderate thermal management improvements. 

A study has been performed to determine the most economic launch option and optimal HD-SPS sizing based on near-term launch vehicles. The SpaceX Falcon Heavy launch vehicle was chosen to launch utility-scale mid-term HD-SPS. While essentially offering the same payload volume as a Falcon $9$ vehicle, the payload mass launched on a Falcon Heavy to LEO is quoted as about $53$~tons~\cite{spacexfalcon9}. Assuming that $10\%$ of the total payload volume as well as payload mass is needed for packaging material, and that $70\%$ of the remaining volume and mass can be used for the integration of sandwich modules, about $79$ sandwich modules with the improved specifications described above can be launched on each Falcon Heavy vehicle, capable of $630~kW$ nameplate RF power at $3$-sun concentration.

Since the HD-SSP concept does not rely on pre-existing in-space assembly infrastructure, also the mid-term GLPO HD-SPS will self-deploy once final orbit is reached.

\subsection{Orbit Transfer}

The HD-SSP concept also involves independence of in-space orbital transfer infrastructure. To reach the final GLP orbit after launch from LEO, each mid-term HD-SPS will feature on-board electric propulsion thrusters. The necessary $\Delta v$ to reach GEO from LEO is about $4.7$~km/s. Orbital transfer can be performed via a low-thrust spiral transfer orbit. Assuming a specific power of $I_{SP}=3000$~s for electric propulsion, it can be deduced from the rocket equation that a $20$~ton SPS will use about $3.4$~tons of propellant~\cite{Ianperscorrespondence}. While xenon propellant could represent a significant cost contribution on the order of millions of USD, argon would provide very low-cost fuel. The required amount of $3.4$~tons of argon or xenon propellant can be stored in its liquid phase in a storage vessel with a capacity of about $1$~m$^{3}$. 

To minimize damage of the sandwich modules and reflector arrays during transit, the HD-SPS can be left undeployed in the compact payload integration configuration. That way the elements of the satellite expose minimal cross section to damage from orbital debris, micrometeoroids and radiation trapped in the Van Allen belts. The satellite can be fully deployed after reaching GLPO to start producing RF power. The payload configuration stack of sandwich modules could be surrounded by an additional outside layer of protective material, which is discarded at arrival at GLPO or could be used for other purposes, such as radiator material. 

Power to the thrusters, which could be located at the bottom of the payload volume, can be supplied by additional flat photovoltaic modules with identical size as the sandwich modules. These additional modules will be deployed first at LEO, providing sufficient power to the electric propulsion system. With about $11$~m$^{2}$ active photovoltaic area each and $20\%$ efficiency, $10$ photovoltaic modules circularly surrounding the payload stack could provide about $31$~kW of power, sufficient to power one or several electric thrusters. 



\subsection{Thinned Array Curse}

In order to reach utility-scale power levels at the ground rectenna, spatial combination of beams from a multitude of HD-SPS is required. The electromagnetic superposition of power beams from multiple, circularly distributed sources results in a more focused and directed power beam which requires a smaller rectenna area to be properly intercepted. However, the main lobe beam power level decreases and sidelobe levels increase by a fraction proportional to the area between single radiators relative to the total area occupied by the radiator constellation, which is described by the concept of the 'thinned-array-curse'~\cite{wikipediathinnedarraycurse}. 
This makes power beam combination an economically challenging concept for SPS, since only a fraction of the launched RF power can be received on the ground and more SPS have to be launched to achieve a desired amount of grid power. 

The design concept of HD-SSP requires the SPS to receive sunlight on their photovoltaic surfaces from reflector arrays which are properly adjusted to the respective position of the Sun during one orbit in GLPO. For instance, at local midnight at the ground station, sunlight has to be redirected by $180$~degrees to reach the photovoltaic sandwich surfaces facing away from the Earth. If all HD-SPS in a constellation of $N$ satellites are distributed over a circular area at the same radial distance to the Earth, three-sun concentration on the photovoltaic surfaces requires a spacing between neighboring satellites of at least three times the area of a single SPS to not block sunlight before it reaches the reflector arrays. This requires the launch of $4 \times N$ SPS to achieve a power level on the ground corresponding to $N$ SPS (albeit over a smaller rectenna area). To remedy these difficulties with HD-SPS, a 'staggered' HD-SPS constellation is explored in the following. 

\subsection{Staggered HD-SPS GLPO Constellations}

\begin{figure}[h]
	\begin{center}
		\includegraphics[width=0.4\textwidth]{{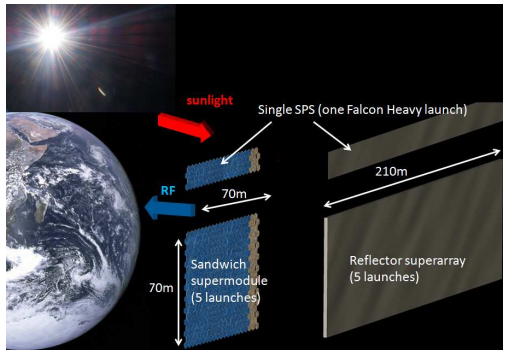}}
		\caption{\label{fig:supermodules} 
		Single launch GLPO HD-SPS and sandwich supermodules and reflector superarrays formed from $5$ HD-SPS are shown. Images of Sun and Earth from Wikipedia.}  
	\end{center}
\end{figure}

To deliver utility-scale power, the sandwich modules of one GLPO HD-SPS satellite are configured in a rectangular shape with a long edge length of about $70$m. The corresponding reflector array has the same rectangular shape as well, only with a length of $210$~m to be able to provide $3$-sun concentration to the sandwich modules. Five HD-SPS are connected to form a HD-SPS 'supermodule' sandwich platform of square shape with $70$~m side length and reflector 'superarrays' of $70$~m short edge and $210$~m long edge length. Depictions of single launch GLPO HD-SPS, and supermodules and superarrays are given in Figure~\ref{fig:supermodules}. 

The sandwich supermodules are arranged in a continuous, circular shape as seen from Earth to avoid losses due to the thinned array curse. Both the sandwich supermodules and reflector superarrays are physically connected along the north-south direction. About $20$ strings of sandwich supermodules are required to form a circular area capable of $1$~GW RF nameplate power, corresponding to about $1600$ HD-SPS launched by the same number of Falcon Heavy vehicles. The radius of such a constellation is about $710$~m, and the ground rectenna diameter to capture $>95\%$ of the RF power generated can be calculated from Equation~\ref{eq:taueq} to be about $7.8$~km. 
 
\begin{figure}[h]
	\begin{center}
		\includegraphics[width=0.4\textwidth]{{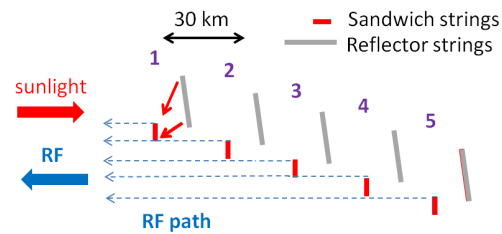}}
		\caption{\label{fig:glpoconstellation_nsview} 
		The first five north-south strings of HD-SPS of an exemplary GLPO constellation with three-sun concentration are depicted at local midnight at the receiver station, looking along the north-south direction along each satellite string. Blue dashed lines illustrate the areas of $>95\%$ of the total beam power from each sandwich supermodule string.}  
	\end{center}
\end{figure}

To maintain a circular shape as seen from Earth but to enable concentrated light redirection from the reflector superarrays to the sandwich supermodules, the sandwich HD-SPS strings are positioned at different orbital altitudes. As an appropriate radial distance between neighboring strings, the length of the umbra of a north-south string of supermodules can be chosen. Due to the non-zero diameter of the Sun as seen from the Earth, light emitted from opposite sides of the Sun at ground receiver local time noon and midnight meets at a distance of about $30$~km behind each GLPO HD-SPS. For the next-neighbor string along the direction from the Sun to the Earth, the shadow of the previous string will cover some but not all of the Sun, allowing for continued albeit reduced power production at ground receiver midnight and noon hours. At all other times, the intensity reduction and Sun obstruction for strings located behind other strings will be less than at midnight and noon hours. Around $6$~am and $6$~pm ground receiver local time, full $3$-sun conentration light intensities incide on the sandwich modules in each string, as the shadows of neighboring strings do not pass over any reflector arrays. Figure~\ref{fig:glpoconstellation_nsview} shows a north-south (projecting) view of the first five supermodule and superarray strings, which are closest to Earth, of a $1$~GW RF constellation. Three-sun concentration sunlight is collected by the reflector superarray strings, which are located relative to the sandwich modules at the side of each sandwich supermodule string such as to not block the RF beam paths of neighboring strings. 

An upper limit of beam power intercepted of sandwich supermodule strings (causing additional heat load) from all other neighboring strings can be given by $21\%$ of the total incident power on each supermodule. This number is calculated from comparing the size of the $95\%$ power footprint of the beams of neighboring $70$~m $\times$ $70$~m supermodules at the location of their next-neighbor string $30$~km away with the $70$~m on orbit displacement between neighboring strings. Tolerating an at maximum $21\%$ additional heat load on sandwich modules is assumed to be a technologically solvable challenge.

\subsection{Estimated Cost of Utility-Scale HD-SPS}

Analogously to Section~\ref{sec:leodemocost}, a learning curve approach is used to estimate the cost for a utility-scale HD-SPS constellation with the parameters specified in Section~\ref{sec:glpoconstellation}. For the large numbers of sandwich modules and subsystems required for utility-scale SPS systems, a parametrization of learning curves similar as outlined in Ref.~\cite{mankinsniac} is employed. The production cost of a unit drops by $34\%$ at every doubling of the production volume. The total cost for the production of $N$ units is calculated by integrating the corresponding cost curve up to the respective total number of modules. For launch costs, a similar approach is applied, however with a learning curve improvement smaller by $10\%$ compared to the improvement rate used in Ref.~\cite{mankinsniac}. This amounts to a reduction of cost by $27\%$ each time the number of launches is doubled.

For launching a constellation of $1600$ HD-SPS to GLPO with Falcon Heavy vehicles, and assuming similar cost fractions as used in the estimation of LEO demonstrator costs, a utility scale HD-SPS constellation with three-sun concentration is estimated to be developed, produced and launched for about $15$~billion USD. 

Locating the rectenna for the first HD-SPS constellation in New England and selling the power for $14$~cents/kWh, so cheaper than the average retail price for all end-use sectors~\cite{eia.gov}, would allow to amortize the production and launch cost in about $11$~years starting from full-scale operation, or earlier if electricity is sold already before reaching nameplate capacity in GLPO.

Further extrapolation of cost for decades after launches for a GLPO HD-SPS constellation have begun is beyond the scope of this study and also considered of limited helpfulness. Technological development over the scale of decades will probably achieve advances which would necessitate modification or substitution of HD-SPS with more optimized SSP system architectures.


\section{Challenges for HD-SPS Staggered Design}

Further studies have to be performed if a staggered design can in fact minimize losses due to the thinned array curse. A circular shape as viewed from Earth would not incur thinned array losses, however the sandwich supermodule strings would be at significantly different orbital altitudes. The behavior of the combined beam profile of all strings would need to be studied.

Another challenge for the staggered design is to maintain the circular constellation shape as seen from Earth. Due to the large difference in orbital altitudes of about $580$~km between the strings closest and furthest from Earth, their orbital periods also differ by about $1700$~seconds. To achieve the same orbital periods and to maintain the overall circular shape as seen from Earth, the outermost string would need to be sped up by about $64$~m/s beyond its orbital equilibrium velocity. This would require a continuous centripetal acceleration of about $490$~N to maintain orbit. Chemical propulsion units would be needed for this task, which would introduce a large amount of propellant to be launched as well considering the expected operating lifetime of SPS on the order of $10$~years.

\section{Comparison of HD-SPS to Other SPS Designs}

The main design principle for HD-SPS is to avoid relying on to-be-developed in-space infrastructure. Space infrastructure development will require funding as well as manpower and time in addition to the resources already necessary to build SPS themselves. HD-SPS could provide a design which allows deploying utility-scale SSP systems within $10$ years of production start (assuming $100$ Falcon Heavy launches per year), without in-orbit assembly or orbital transfer infrastructure. Infrastructure to repair damaged HD-SPS, such as a GLPO repair platform, can be useful. Defective HD-SPS with functional station-keeping systems can be maneuvered out of the constellation into a repair orbit, where robotic or tele-operated maintenance can be performed. New HD-SPS can be continuously resupplied to existing constellations by ongoing launches.

Since each launch delivers a complete and fully functional HD-SPS to its final orbit, power harvesting can start with the first deployed HD-SPS, or at least after a few hundred HD-SPS are delivered to their final orbits. Other designs rely on complete deployment of the full system including mirror arrays and other SPS components before first baseload space solar power transmitted to Earth.

The mostly infrastructure-independent concept for HD-SPS allows the dispatch of arbitrary numbers of HD-SPS to different orbits and other points of interest for space power, such as the Lagrange libration points or orbits around the Moon.

The limited size of a complete HD-SPS avoids yet to be developed very large scale structures, which again reduces time and cost to develop and deploy SSP. All components of an HD-SPS can be produced and integrated with existing technology of reflectors, photovoltaic cells, power management, thermal management and microwave antennas. Mass production is likely to decrease the cost for HD-SPS over the years, as described in Ref.~\cite{mankinsniac}. In addition, real time surveying of the photovoltaic and RF antenna surfaces is more easily accomplished than for km-scale diameter apertures.

Due to the limited size of HD-SPS sandwich supermodules along orbit direction of $70$~m, each sandwich module is in relatively close proximity to the outer surface of the supermodule strings. This makes it possible to include heat transfer systems in the satellite design and potentially also deployable light-weight radiators on the sides of the main platform. As a consequence, the main platform of each HD-SPS could be irradiated with higher concentration factors than in other, similarly structured but much larger platform designs.  


\section{Summary}

A high-level system study of a scalable concept for SSP without relying on in-space assembly or transportation infrastructure is presented. The total cost of two LEO HD-SPS demonstrators and a ground station to receive the transmitted power is estimated in a high-level economic analysis to about $1.9$~billion USD. The concept seems to offer scalability to utility-sized satellite constellations in GLPO. However, further system design research is required, in addition to the need for further development of spatial microwave beam combination. Total cost for a $1$~GW utility-scale HD-SPS constellation in GLPO was estimated to about $15$~billion USD, using a similar learning curve approach as Ref.~\cite{mankinsniac}.

\section{Acknowledgments}
I would like to thank Mr. Ian McNally from the University of Glasgow, Scotland, UK, Mr. Paul Jaffe from the Naval Research Laboratory and Prof. Greg Durgin from the Georgia Institute of Technology, both USA, for helpful discussions and suggestions.


\end{document}